\documentclass[conference]{IEEEtran}
\usepackage{etoolbox}
\IEEEoverridecommandlockouts

\usepackage{cite}
\usepackage{amsmath,amssymb,amsfonts}
\usepackage{algorithmic}
\usepackage{graphicx}
\usepackage{textcomp}

 \usepackage{subcaption}
  \usepackage{graphicx}
  \usepackage{lipsum}

\usepackage{fancyhdr}

\usepackage{xcolor}
\def\BibTeX{{\rm B\kern-.05em{\sc i\kern-.025em b}\kern-.08em
    T\kern-.1667em\lower.7ex\hbox{E}\kern-.125emX}}


\begin{document}

\title{Recognition of Frequencies of Short-Time SSVEP Signals Utilizing an SSCCA-Based Spatio-Spectral Feature Fusion Framework
}

\author{\IEEEauthorblockN{Saif Bashar}
\IEEEauthorblockA{\textit{Department of CSE} \\
\textit{Jamalpur Science \& Technology University}\\
\textit{Jamalpur-2012, Bangladesh}\\
Email: saifbashar2021@gmail.com}
\and
\IEEEauthorblockN{Samia Nasir Nira}
\IEEEauthorblockA{\textit{Department of CSE} \\
\textit{Jamalpur Science \& Technology University}\\
\textit{Jamalpur-2012, Bangladesh}\\
Email: samiyanira9842@gmail.com}
\and
\IEEEauthorblockN{Shabbir Mahmood}
\IEEEauthorblockA{\textit{Department of CSE} \\
\textit{Pabna University of}\\
\textit{Science and Technology}\\
Pabna, Bangladesh\\
Email: shabbir.cse@pust.ac.bd}
\and
\IEEEauthorblockN{Md. Humaun Kabir}
\IEEEauthorblockA{\textit{Department of CSE} \\
\textit{Jamalpur Science \& Technology University}\\
\textit{Jamalpur-2012, Bangladesh}\\
Email: humaun@bsfmstu.ac.bd}
\and
\IEEEauthorblockN{Sujit Roy}
\IEEEauthorblockA{\textit{Department of CSE} \\
\textit{Jamalpur Science \& Technology University}\\
\textit{Jamalpur-2012, Bangladesh}\\
Email: sujit@bsfmstu.ac.bd}
\and
\IEEEauthorblockN{Iffat Farhana}
\IEEEauthorblockA{\textit{Department of CSE} \\
\textit{Varendra University}\\
Rajshahi, Bangladesh \\
Email: iffat@vu.edu.bd}
}
\maketitle
\begin{abstract}
A brain-computer interface (BCI) facilitates direct communication between the brain and external equipment through EEG, which is preferred for its superior temporal resolution. Among EEG techniques, the steady-state visual evoked potential (SSVEP) is favored due to its robust signal-to-noise ratio, minimal training demands, and elevated information transmission rate. Frequency detection in SSVEP-based brain-computer interfaces commonly employs canonical correlation analysis (CCA). SSCCA (spatio-spectral canonical correlation analysis) augments CCA by refining spatial filtering. This paper presents a
multistage feature fusion methodology for short-duration SSVEP frequency identification, employing SSCCA with template signals derived via leave-one-out cross-validation (LOOCV). A filterbank generates bandpass filters for stimulus frequencies and their harmonics, whereas SSCCA calculates correlation coefficients between subbands and templates. Two phases of non-linear weighting amalgamate these coefficients to discern the target stimulus. This multistage methodology surpasses traditional techniques, attaining a accuracy of 94.5\%.
\end{abstract}
\section{Introduction}
A Brain-Computer Interface (BCI) transforms neurophysiological data into digital commands, facilitating communication with external devices without the involvement of peripheral nerves or muscles, hence providing opportunities for individuals with significant motor impairments \cite{wolpaw2002brain}. Electroencephalography (EEG) is extensively utilized in brain-computer interfaces (BCIs) owing to its mobility, cost-effectiveness, and superior temporal resolution. SSVEP-based BCIs are favored due to their elevated signal-to-noise ratio, minimal training demands, and robust information transfer rate, utilizing visual stimuli at distinct frequencies identified by frequency recognition algorithms. Although CCA is frequently employed for frequency recognition, its accuracy limits and computational expenses have prompted the advancement of enhanced methodologies \cite{lin2006frequency}.

\section{Problem Statement}
CCA-based algorithms are efficient for SSVEP-based BCIs but encounter constraints such as inferior accuracy and elevated computing expenses due to dependence on predetermined reference signals devoid of EEG characteristics \cite{zhang2014frequency}. Advanced techniques such as MsetCCA and MCM enhance reference signals but encounter difficulties with noise extraction and insufficient utilization of SSVEP harmonics \cite{jiao2018novel}. Recent methodologies such as FBMSI and SSCCA enhance accuracy and decrease computational expenses, although frequently eliminate essential discriminative information, hence affecting performance \cite{cherloo2022spatio}.
\section{Literature Review}
CCA, as presented by Lin et al. \cite{lin2006frequency}, use sinusoidal reference signals for frequency identification; however, it necessitates calibration and does not incorporate actual EEG characteristics. MsetCCA boosts reference signals utilizing training data \cite{zhang2014frequency}, whereas MCM improves accuracy by diminishing noise and amplifying correlations \cite{jiao2018novel}. FBMSI enhances harmonic exploitation for improved SSVEP recognition \cite{qin2021filter}, while SSCCA simplifies analysis through a singular projection vector, hence augmenting efficiency and accuracy \cite{guccluturk2016liking}. Nevertheless, conventional CCA and CORRCA preserve solely the maximum correlation coefficient, resulting in information loss \cite{mahmood2021frequency}. To resolve this, filter bank analysis and feature fusion frameworks amalgamate subband features for enhanced frequency recognition \cite{mishuhina2018feature}, hence augmenting accuracy and information transfer rate for SSVEP-based brain-computer interfaces.

\section{Dataset Description}
The publicly accessible dataset from the Swartz Center for Computational Neuroscience at UC San Diego derives from an offline SSVEP-based BCI experiment with 10 individuals (9 males, 1 female; average age 28 years). Five individuals possessed prior expertise with SSVEP-BCI, whereas the others were novices. Written consent was acquired \cite{nakanishi2015comparison}. The 12-class SSVEP dataset was acquired via a Biosemi Active Two EEG equipment equipped with eight occipital electrodes. Stimuli were encoded using frequencies ranging from 9.25 Hz to 14.75 Hz in increments of 0.5 Hz, and phases of 0, 0.5$\pi$, $\pi$, and 1.5$\pi$. EEG data were initially collected at 2048 Hz and subsequently downsampled to 256 Hz. Each participant underwent 15 trials, each consisting of 12 randomly sorted targets shown in a 4×3 grid (6×6 cm) for 4 s, with a 135 ms visual delay.
\section{METHODOLOGY}
\subsection{SPATIO-SPECTRAL CANONICAL CORRELATION ANALYSIS (SSCCA)}
The spatio-spectral canonical correlation analysis (SSCCA) method, originating from the common spatio-spectral patterns (CSSP) technique, improves spatial and frequency information by delaying EEG input.

Canonical correlation analysis (CCA) maximizes the correlation between two sets of variables, \(Z\) and \(Y\), utilizing weight vectors \(W\) and \(V\).In this context, $Z$ (of dimensions $N_c \times N_s$) and $Y$ (of dimensions $N_c \times N_s$) are matrices comprising real values, with canonical variables defined as $H_{z} = W^{T}Z$ and $H_{y} = V^{T}Y$. The spatial filters $W$ and $V$ are engineered to synchronize the EEG and reference signals, hence optimizing their correlation $\rho$ which may be computed using Eq. \ref{eqn1}.
\begin{center}

\begin{equation}\label{eqn1}
\rho = arg_{H}max_{v}\frac{corr(H_{z}, H_{y})}{\sqrt{var(H_{z})} \sqrt{var(H_{y})}} 
\end{equation}

\begin{equation}
\label{eqn2}
\rho = arg_{H}max_{v}\frac{W^{T}ZY^{T}V}{\sqrt{W^{T}ZZ^{T}W} \sqrt{V^{T}YY^{T}V}}  
\end{equation}
\end{center}

The variables $N_s$ and $N_c$ denote the signal length and the quantity of EEG channels, respectively. $Z$ represents the multi-channel EEG signal, while $Y$ is the reference signal. Eq. \ref{eqn3} and \ref{eqn4} delineates the canonical variables comprehensively.

\begin{equation}
\label{eqn3}
Z=
\begin{bmatrix}
z_{1} \\
. \\
. \\
z_{N_c}
\end{bmatrix}
,W= \begin{bmatrix}
    w_{1}, w_{2},..w_{N_c}
\end{bmatrix}^{T}
Y= \begin{bmatrix}
    y_{1} \\
    .\\
    .\\
    y_{N_c}
\end{bmatrix}\\
\end{equation}
\begin{equation}
,V = \begin{bmatrix}
    v_{1}, v_{2} .. v_{N_c}
\end{bmatrix}^{T}\\ 
\nonumber
\end{equation}
\begin{equation}
\label{eqn4}
    H_{z} = w_{1}z_{1} + w_{2}z_{2} +.. + w_{N_c}z_{N_c}  \\
    ,H_{y} = v_{1}y_{1} + v_{2}y_{2} + ...+v_{N_c}y_{N_c}
\end{equation}

To ascertain the SSVEP frequency using CCA, reference signals for each visual stimulus, represented as $[Y_{f^{1}},Y_{f^{2}}...,Y_{f^{k}}]$, are generated. In this context, $k$ represents the quantity of stimuli, while $f^{k}$ denotes the frequency of the $k^{th}$ stimulus. CCA calculates correlation coefficients $[\rho_{1},\rho_{2}...\rho_{r}]$, where $r$ represents the minimum rank of the matrices $Z$ (EEG signals) and $Y$ (reference signals). The correlation results indicate a robust relationship between EEG and reference signals for each stimulus, especially between the SSVEP frequency and the reference. Time-delay embedding improves CCA efficacy by integrating both original and delayed EEG signals as inputs.Thus, Eq. \ref{eqn3} is modified in the following manner
\\
\begin{equation}\label{eqn5}
\hat{Z} = \begin{bmatrix}
Z\\
\delta _ \tau Z
\end{bmatrix},
\hat{H}_z = W_0^TZ + W_\tau^TZ(\delta _\tau Z) 
\end{equation}
The expanded form of the Eq. \ref{eqn5} and the canonical variable will be in 
the form of Eq. \ref{eqn6}:
\begin{equation}\label{eqn6}
    Z = \begin{bmatrix}
    z_1\\
    .\\
    .\\
    z_{N_c}\\
    \delta _ \tau z_1\\
    .\\
    .\\
    \delta _ \tau z_{N_c}
\end{bmatrix},
W = \begin{bmatrix}
    w_1^0,  w_2^0 ,.. w_p^0,w_1^\tau,  w_2^\tau ,.. w_{N_c}^\tau
\end{bmatrix}^T,
\nonumber
\end{equation}
\begin{equation}
Y = \begin{bmatrix}
    y_1\\
    .\\
    .\\
    y_{N_c}
\end{bmatrix},
    V = \begin{bmatrix}
    v_1,  v_2 ,.. v_{N_c}
\end{bmatrix}^T\\
\end{equation}

The canonical variables of the embedded EEG signal and the reference signal will be expressed in the form of the Eq. \ref{eqn7}.
\begin{equation}\label{eqn7}
\begin{split}
    \hat{W_z} = W_0^TZ(n) + W_\tau^TZ(n-\tau) = v_1y_1 +  .. + v_{N_c}y_{N_c} = V^TY
\end{split}
\end{equation}

Eq. \ref{eqn7} is similar to FIR filter, therefore, when calculating the canonical 
variables $\hat{H}_x$, The FIR frequency filter is utilized in conjunction with spatial filtering, whereas time delay embedding facilitates concurrent frequency and spatial optimization during the correlation computation. The correlation coefficient ($\rho$) shown in Eq. \ref{eqn1} and Eq. \ref{eqn2}.

We may utilize the SSCCA to identify the frequency of the SSVEPs.
Using the \ref{eqn2} formula $N_c$ numbers of coefficients $\rho= [\rho_1, \rho_2, ....\rho_{Nc}]$. The maximum correlation coefficient is employed for frequency recognition, with the associated template signal frequency determined as that of the test signal.
, 
\begin{equation}
    f_{test} = max_i(\rho_i)
\end{equation}
\subsection{PROPOSED METHOD}
\begin{figure}[ht]
    \centering
    \includegraphics[width=0.5\textwidth]{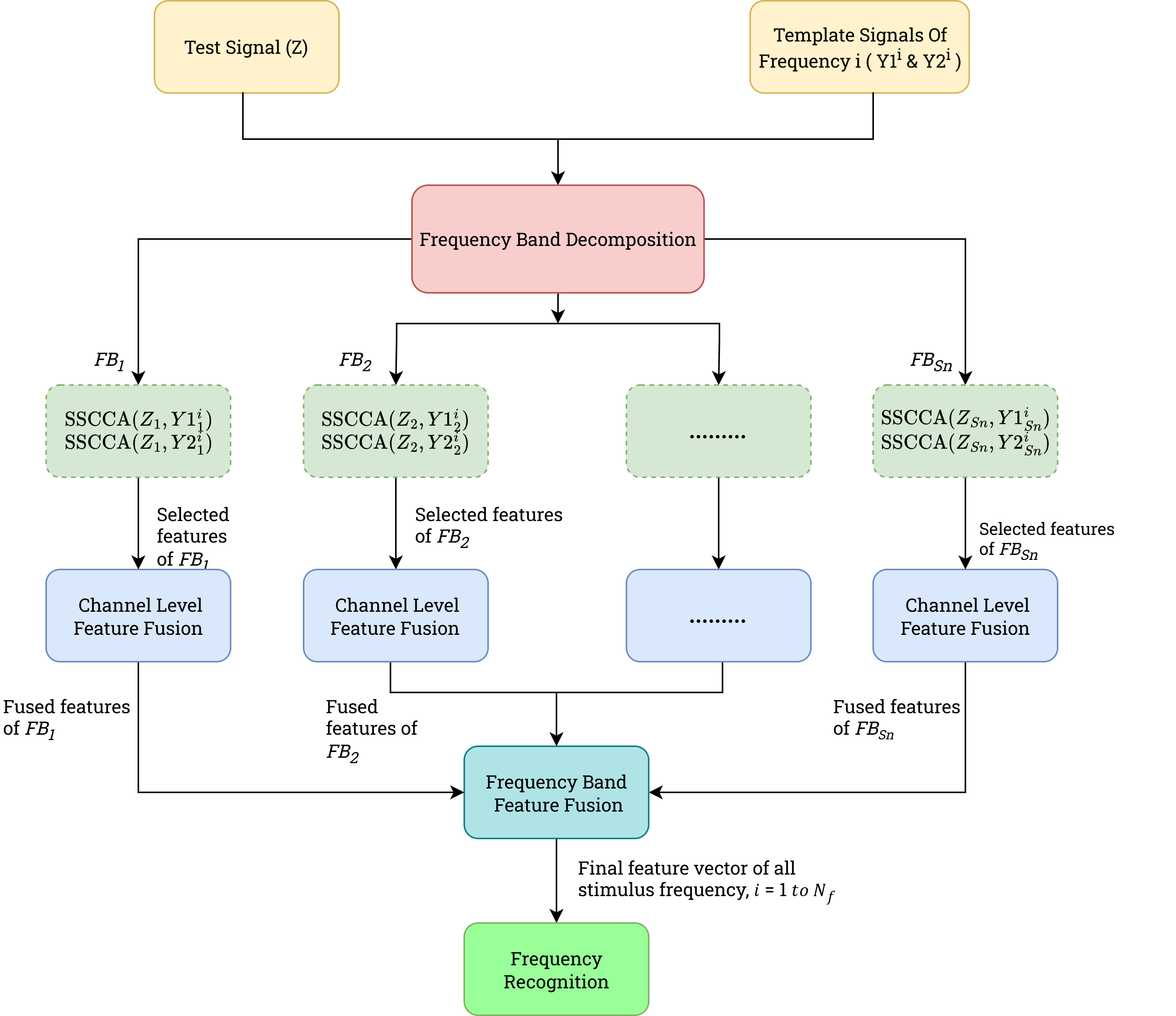}
    \caption{Block Diagram of the Proposed Methodology}
\end{figure}
This study presents a methodology based on SSCCA for the integration of spatio-spectral features to detect SSVEP frequencies. The framework comprises three stages: (1) Subband decomposition; (2) Feature fusion; and (3) Frequency recognition are elaborated upon in the subsequent subsections.
\subsubsection{SUBBAND DECOMPOSITION}
Initially, bandpass filters are employed to partition data into discrete subbands. The dataset comprises four elements: 8 channels, 1041 sample points, 15 trials, and 12 stimuli, resulting in a $N_c \times N_{s} \times N_{t} \times N_{f}$ array. The stimulus frequency varies from 9.25 Hz to 14.75 Hz, with an interval of 0.5 Hz. The utilization of filterbank technology enhances BCI performance. Each trigger undergoes processing using a Chebyshev Type I filter ($12^{th}$ order, 3dB stopband attenuation) to generate Sn subbands, each characterized by unique harmonic frequencies and a shared cutoff of 80 Hz.The lower cutoff frequency of the $m^{th}$ bandpass filter is denoted as $m \times f_0$, where $f_0$ signifies the initial stimulus frequency and $m=1,2,.,Sn(Sn = 5)$.

$Z$ denotes the test signal, whereas $Y1^i$ and $Y2^i$ signify the template signals at the i-th frequency in $\ \mathbb{R}^{N_c \times N_s}$. $Y1^i$ represents the mean of the first segment (trial 1 to trial 7) of the training trials, while $Y2^i$ denotes the mean of the latter segment (trail 8 to trial 14). Subsequent to filtering, the test signal is partitioned into frequency subbands $Z_{1},Z_{2} ... Z_{Sn}$. Correspondingly, the template signal is partitioned into subbands $Y1^{i}_{1},Y1^{i}_{2},....Y1_{Sn}^i$ and $Y2^{i}_{1},Y2^{i}_{2},....Y2_{Sn}^i$.

\subsubsection{FEATURE FUSION}
During the frequency band feature fusion, features are recovered using SSCCA to assess the correlation between each pair of subband signals from the test and template signals. The quantity of features is equivalent to the quantity of channels, $N_c$

For each subband, two coefficient vectors are computed utilizing SSCCA: one derived from the test signal $Z_{m}$ and the first template signal $Y1_m^i$ , and the other from $Z_m$ and the second template signal $Y2_m^i$. The index $m$ denotes the subbands ($m = 1, 2, ..., Sn$). Each pair produces $N_c$ coefficients, which are consolidated into 2 × $N_c$ coefficients. Upon arranging them in descending order, the first $N_c$ coefficients are chosen as the feature vector. In the initial phase of feature fusion, individual coefficients from SSCCA are amalgamated, with the template signal established as the average of all training trials (trial 1 to trial 14).

\begin{figure}[ht]
  \begin{subfigure}{0.45\columnwidth}
  \includegraphics[width=\textwidth]{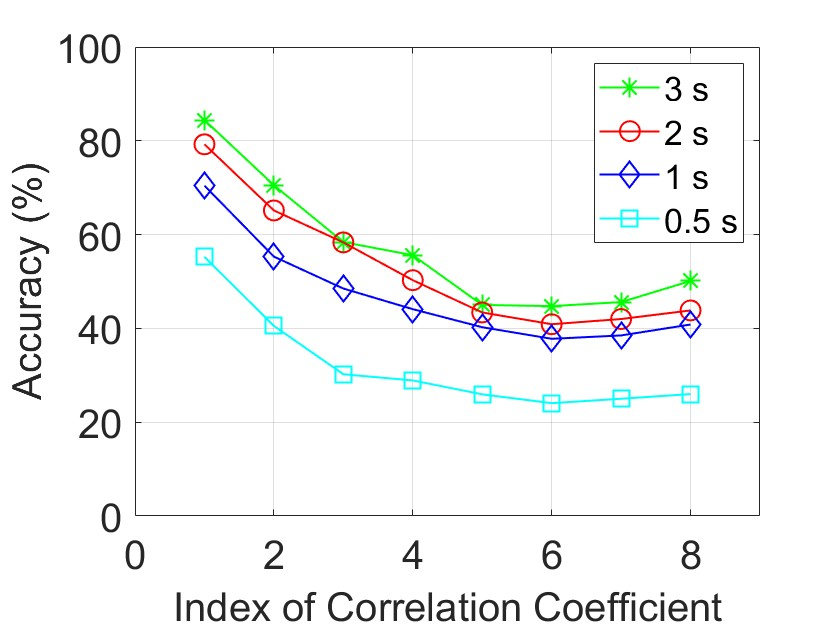}
  
  \caption{Accuracy of individual coefficient}
  \label{first_fig_ff1}
  \end{subfigure}
  \hfill
  \begin{subfigure}{0.45\columnwidth}
  \includegraphics[width=\textwidth]{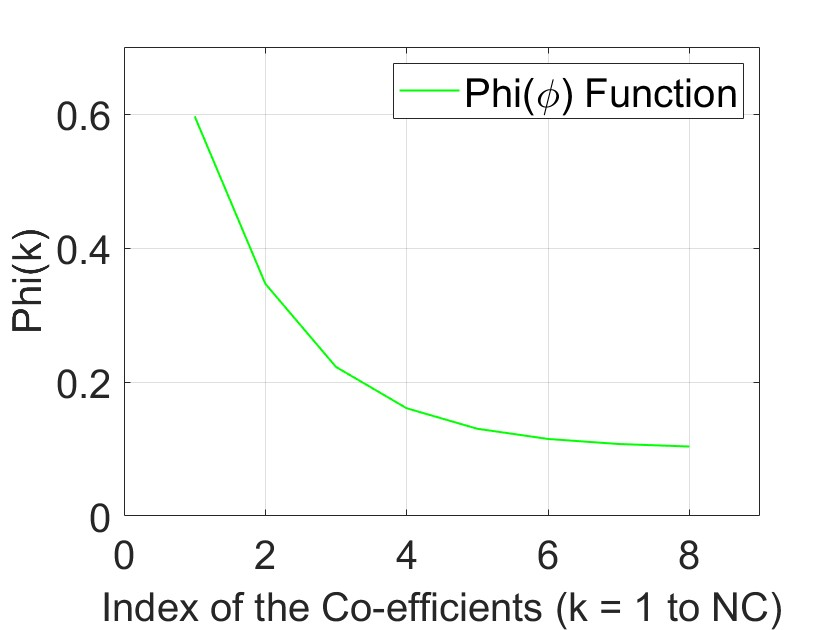}
  
  \caption{Non-linear weighting function}
  \label{first_fig_ff2}
  \end{subfigure} 
  
  \caption{Accuracy of individual coefficient and Non-linear weighting function in channel level feature fusion.}
  \end{figure}

Fig. \ref{first_fig_ff1} illustrates the correlation coefficient index in relation to average classification accuracy among individuals for four temporal intervals (0.25s, 0.50s, 0.75s, and 1.00s), exhibiting a non-linear decline in accuracy.

The non-linear weighting function $\phi$ depicted in Fig. \ref{first_fig_ff2} improves system performance through the integration of coefficients.

\begin{figure}[ht]
  \begin{subfigure}{0.45\columnwidth}
  \includegraphics[width=\textwidth]{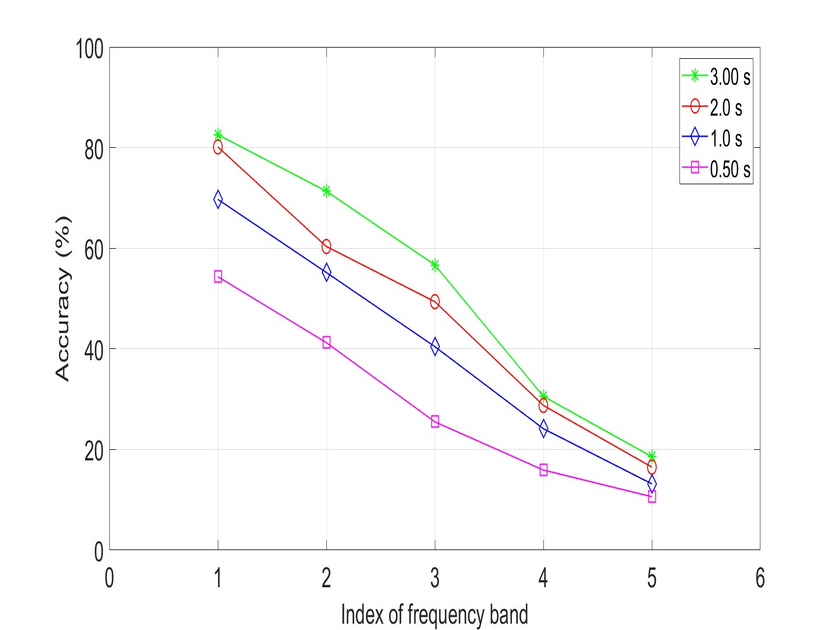}
  
  \caption{Accuracy of individual coefficient}
  \label{second_fig_ff3}
  \end{subfigure}
  \hfill
  \begin{subfigure}{0.45\columnwidth}
  \includegraphics[width=\textwidth]{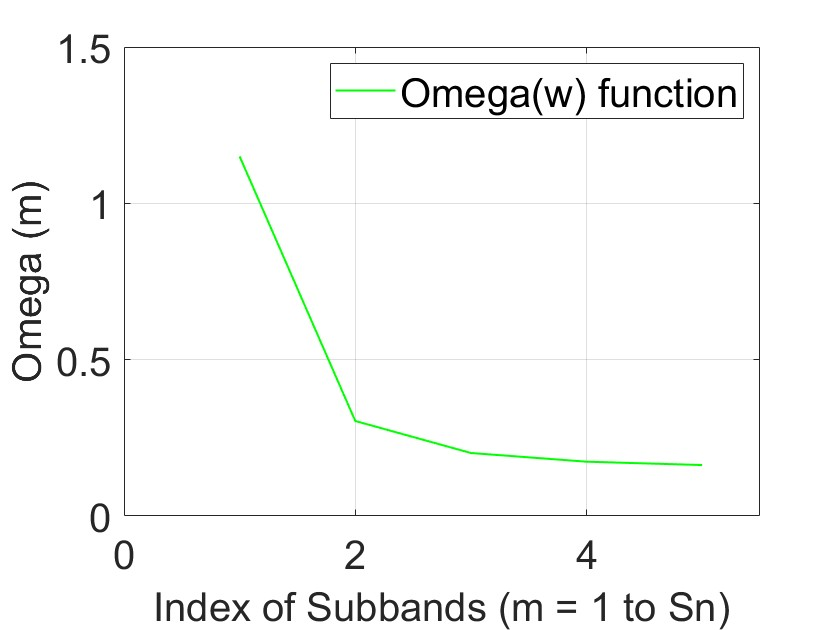}
  
  \caption{Non-linear weighting function(Omega)}
  \label{second_fig_ff4}
  \end{subfigure} 
  
  \caption{Accuracy of individual coefficient and Non-linear weighting function in frequency band feature fusion.}
  \end{figure}

This study utilizes a channel level feature fusion methodology, amalgamating coefficients \( \phi_1, \phi_2, \ldots, \phi_{N_c} \) through a non-linear weighting function \( \phi_k = e^{-a_1 k} + b_1 \), where \( k = 1, 2, \ldots, N_{c} \), \( a_1 = 0.6 \), and \( b_1 = 0.2 \). Grid search optimizes these parameters. The correlation coefficients, obtained using SSCCA, signify distinct occipital brain regions and are amalgamated to create the \( m^{th} \) subband feature.
Each subband coefficient encompasses essential information for identifying stimulus frequencies. Fig. \ref{second_fig_ff3} depicts the correlation between frequency band indices and mean classification accuracy throughout four temporal periods (0.25s, 0.50s, 0.75s, and 1.00s), demonstrating a persistent non-linear decrease in accuracy.

Fig. \ref{second_fig_ff4} illustrates the non-linear weighting function (\(\Omega\)) employed in the frequency band feature fusion to improve system performance through the amalgamation of coefficients.


The combined feature for the test signal \( Z \) is produced using the template signals \( Y1^{i} \) and \( Y2^{i} \) at the \( i \)-th frequency (\( i = 1, 2, \ldots, N_f \)) and is calculated as:
\begin{equation} \psi_i = \sum_{m=1}^{S} w_m \cdot \delta_m^i \label{eq:Li_equation} \end{equation}
The non-linear weighting function is defined as \( w_m = m^{-a_2} + b_2 \), with \( m = 1, 2, \ldots, Sn \), \( a_2 = 2.0 \), and \( b_2 = 0.25 \). The settings are tuned by grid search. This process generates fused features \( \psi_i \) for each frequency \( i \).

\subsubsection{FREQUENCY RECOGNITION}

The concluding process involves determining the stimulus frequency. The attributes acquired from the preceding channel level and frequency band feature fusion are utilized for each stimulus frequency,$\psi_1...\psi_2...\psi_{N_{f}}$. The equation calculates the frequency $F_z$ of the test signal $Z$.
\begin{equation}
    F_z = max_f\psi_i(f), i = 1,2,...N_{f}
\end{equation}
\section{Result \& Performance Analysis}
\begin{figure}[ht]
    \centering
    \includegraphics[width =  0.35\textwidth]{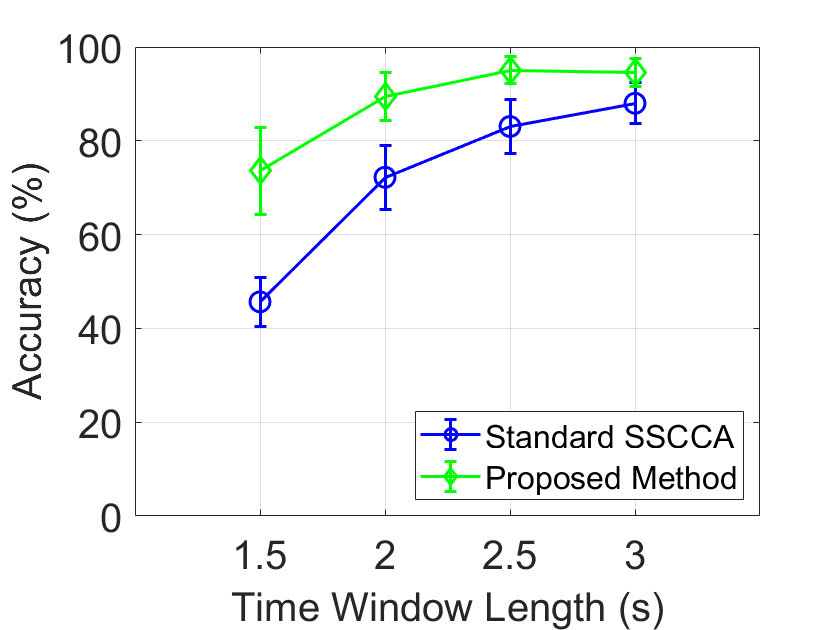}
    \caption{ The average frequency recognition accuracy of SSCCA and the suggested technique among individuals at different time intervals. }
    \label{fig:final_proposed}
\end{figure}
The efficacy of the proposed method was assessed using precision as the evaluation parameter. The average accuracy during a 3s interval was compared to the traditional SSCCA approach, as illustrated in Fig. \ref{fig:final_proposed} The findings, accompanied by error bars representing standard errors, illustrate that the suggested SSCCA spatial filtering regularly surpasses the conventional SSCCA, with a peak accuracy of 94.5\% with a 3s time window and a 1s delay.
\begin{table}[ht]
\centering
\caption{Performance Comparison of Different Models}
\label{tab:comparison}
\begin{tabular}{|l|c|}
\hline
\textbf{Model} & \textbf{Accuracy} \\ \hline
Standard CCA\cite{nakanishi2015comparison} & 55\% \\ \hline
MwayCCA\cite{nakanishi2015comparison} & 68.39\% \\ \hline
L1-MCCA\cite{nakanishi2015comparison} & 70.28\% \\ \hline
MsetCCA\cite{waytowich2018compact} & 73.61\% \\ \hline
UD-C-CNN\cite{ravi2020comparing}& 92.33\% \\ \hline
CORRCA \cite{waytowich2018compact}& 81.44\% \\ \hline
HSCA\cite{miao2021hybrid}& 88\% \\ \hline
\textbf{Proposed Method}& \textbf{94.5\%} \\ \hline    
\end{tabular}
\end{table}

The suggested model attains an impressive accuracy of 94.5\% (Table \ref{tab:comparison}), illustrating its reliability and efficacy in SSVEP frequency recognition, as well as its potential to improve the performance of real BCI systems.
\section{Conclusion}
This study introduces a multistage feature fusion framework utilizing SSCCA-based spatial filtering for precise SSVEP frequency detection in brain-computer interfaces. Although it enhances accuracy and dependability, its disadvantages encompass dependence on training data, manual analysis, constrained stimuli, and reliance on grid search for parameter optimization. Future endeavors will tackle these issues by augmenting datasets, incorporating recordings, diversifying stimulation frequencies, and creating cost-effective, real-time systems to improve the accessibility and practicality of SSVEP-based BCIs.
\bibliographystyle{IEEEtran}
\bibliography{references}
\end{document}